\documentclass[10pt, conference, compsocconf]{IEEEtran}
\usepackage{graphicx}
\begin{document}
\title{Predicting the Future With Social Media}

\author{\IEEEauthorblockN{Sitaram Asur}
\IEEEauthorblockA{Social Computing Lab\\
HP Labs\\
Palo Alto, California \\
Email: sitaram.asur@hp.com}
\and
\IEEEauthorblockN{Bernardo A. Huberman}
\IEEEauthorblockA{Social Computing Lab\\
HP Labs\\
Palo Alto, California \\
Email: bernardo.huberman@hp.com}}

\maketitle
\begin{abstract}
In recent years, social media has become ubiquitous
and important for social networking and content sharing. And yet, the content that is generated from these websites remains largely untapped.
In this paper, we demonstrate how social media content can be used to predict real-world
outcomes. In particular, we use the chatter from Twitter.com to forecast box-office
revenues for movies. We  show that a simple model built from the rate at which tweets are created about particular
topics can outperform market-based predictors. We further demonstrate
how sentiments extracted from Twitter can be further utilized to
improve the forecasting power of social media.
\end{abstract}

\section{Introduction}

Social media has exploded as a category of online discourse where people create content, share it, bookmark it and network at a prodigious rate. Examples include Facebook, MySpace, Digg, Twitter and JISC listservs on the academic side. Because of its ease of use, speed and reach, social media is fast changing the public discourse in society and setting trends and agendas in topics that range from the environment and politics to technology and the entertainment industry.

Since social media can also be construed as a form of collective wisdom, we decided to investigate its power at predicting real-world outcomes. Surprisingly, we discovered that the chatter of a community can indeed be used to make quantitative predictions that outperform those of artificial markets. These information markets generally involve the trading of state-contingent securities, and if large enough and properly designed, they are usually more accurate than other techniques for extracting diffuse information, such as surveys and opinions polls. Specifically, the prices in these markets have been shown to have strong correlations with observed outcome frequencies, and thus are good indicators of future outcomes~\cite{Pennock2001, Chen2003}.  

In the case of social media, the enormity and high variance of the information that propagates through large user communities presents an interesting opportunity for harnessing that data into a form that allows for specific predictions about particular outcomes, without having to institute market mechanisms. One can also build models to aggregate the opinions of the 
collective population and gain useful insights into their behavior, while predicting future trends. Moreover, gathering 
information on how people converse regarding particular products can be helpful when designing marketing and advertising campaigns~\cite{Leskovec2006, Jansen2009}. 

This paper reports on such a study. Specifically we consider the task of predicting box-office revenues for movies using the chatter from Twitter,  one of the fastest growing 
social networks in the Internet. Twitter~\footnote{http://www.twitter.com}, a micro-blogging network, has experienced a burst of popularity in recent months leading to a huge user-base, consisting of several tens of millions of users who actively participate in the creation and propagation of content.  

We have focused on movies in this study for two main reasons.
\begin{itemize}
\item The topic of movies is of considerable interest among the social media user community, characterized both by large 
number of users discussing movies, as well as
a substantial variance in their opinions.
\item The real-world outcomes can be easily observed from box-office revenue for movies.
\end{itemize}

Our goals in this paper are as follows. First, we assess how buzz and attention is created for different movies and how that changes over time. Movie producers spend a lot of effort and money in publicizing their movies, and have also embraced the Twitter medium for this purpose. We then focus on 
the mechanism of viral marketing and pre-release hype on Twitter, and the role that attention plays in forecasting real-world box-office performance. Our hypothesis is that movies that are well talked about will be well-watched.

Next, we study how sentiments are created, how positive and negative opinions propagate and how they 
influence people. For a bad movie, the initial reviews might be enough to discourage others from watching it, while on the
other hand, it is possible for interest to be generated by positive reviews and opinions over time. For this
purpose, we perform sentiment analysis on the data, using text classifiers to distinguish positively
oriented tweets from negative. 

Our chief conclusions are as follows:
\begin{itemize}
\item We show that social media feeds can be effective indicators of real-world performance.
\item We discovered that the rate at which movie tweets are generated can be used to build a powerful model for predicting movie box-office revenue. Moreover our predictions are consistently better than those produced by an information market such as the Hollywood Stock Exchange, the gold standard in the industry~\cite{Pennock2001}.
\item Our analysis of the sentiment content in the tweets shows that they can improve box-office revenue predictions based on tweet rates only after the movies are released.
\end{itemize}

This paper is organized as follows. Next, we survey recent related work. We then provide a short introduction to Twitter 
and the dataset that we collected. In Section 5, we study how attention and popularity are created and how they evolve. We
then discuss our study on using tweets from Twitter for predicting movie performance. In Section 6, we present our analysis on sentiments and their effects. We conclude in Section 7. We describe our prediction model in a general context in the Appendix.

\section{Related Work}

Although Twitter has been very popular as a web service, there has not been considerable published research on it. Huberman and others~\cite{Huberman2008} studied the social interactions on Twitter to reveal that the driving process for usage is a sparse hidden network underlying the friends and followers, while most of the links represent meaningless interactions. Java et al~\cite{Java2007} investigated community structure and isolated different types of user intentions on Twitter. Jansen and others~\cite{Jansen2009} have examined Twitter as a mechanism for word-of-mouth advertising, and considered particular brands and products while examining the structure of the postings and the change in sentiments. However the authors do not perform any analysis on the predictive aspect of Twitter.

There has been some prior work on analyzing the correlation between blog and review mentions and performance. Gruhl and others~\cite{Gruhl2006} showed how to generate automated queries for mining blogs in order to predict spikes in book sales. And while there has been research on predicting movie sales,  almost all of them have used meta-data information 
on the movies themselves to perform the forecasting, such as the movie’s genre, MPAA rating, running time, release date, the number of screens on
which the movie debuted, and the presence of particular actors or actresses in the cast.
Joshi and others~\cite{Joshi2010} use linear regression from text and metadata features to predict earnings for movies.
Sharda and Delen~\cite{Sharda2006} have treated the prediction problem as a classification problem and used neural networks to classify movies into categories ranging from 'flop' to 'blockbuster'. Apart from the fact that they are predicting ranges over actual numbers, the best accuracy that their model can achieve is fairly low.
Zhang and Skiena~\cite{Zhang2009} have used a news aggregation model along with IMDB data to predict movie box-office numbers. We have shown how our model can generate better results when compared to their method.

\section{Twitter}
Launched on July 13, 2006, Twitter~\footnote{http://www.twitter.com} is an extremely popular online microblogging service. It has a very 
large user base, consisting of 
several millions of users (23M unique users in Jan~\footnote{http://blog.compete.com/2010/02/24/compete-ranks-top-sites-for-january-2010/}). It can be considered a directed social network, where each user has a set of subscribers known as followers.
Each user submits periodic status updates, known as $tweets$, that consist of short messages of maximum size 140 characters. These updates typically consist of personal information about the users, news or links to content such as images, video and articles. 
The posts made by a user are displayed on the user's profile page, as well as shown to his/her followers. It is also possible to send a direct message to another user.
Such messages are preceded by $@user_{id}$ indicating the intended destination. 

A $retweet$ is a post originally made by one user that 
is forwarded by another user. These retweets are a popular means of propagating interesting posts and links 
through the Twitter community.

Twitter has attracted lots of attention from corporations for the immense potential it provides 
for viral marketing. Due to its huge reach, Twitter is increasingly used by news organizations to
filter news updates through the community. 
A number of businesses and organizations are using Twitter
or similar micro-blogging services to advertise products and disseminate information
to stakeholders.

\section{Dataset Characteristics}

The dataset that we used was obtained by crawling hourly feed data from Twitter.com. 
To ensure that we obtained all tweets referring to a movie, we used keywords present 
in the movie title as search arguments. We extracted tweets over frequent intervals using the 
Twitter Search Api~\footnote{http://search.twitter.com/api/}, thereby ensuring we had 
the timestamp, author and tweet text for our analysis.  We extracted 2.89 million tweets referring to 24 different movies released over a period of three months. 

Movies are typically released on Fridays, with the exception of a few which are released on Wednesday. Since an average 
of 2 new movies are released each week, we collected data over a time period of 3 months from November to February 
to have sufficient data to measure predictive behavior. For consistency, we only considered the movies released on a Friday and 
only those in wide release. For movies that were initially in limited release, we began collecting data from the time it became wide. 
For each movie, we define the {\it critical period} as the time from the week before it is released, when the 
promotional campaigns are in full swing, to two weeks after release, when its initial popularity fades and opinions 
from people have been disseminated. 

Some details on the movies chosen and their release dates are provided in Table 1. 
Note that, some movies that were released during the period considered were not used in this study, simply because it was difficult to correctly identify tweets that were relevant to those movies. For instance, for the movie {\it 2012}, it was impractical to segregate tweets talking about the movie, from those referring to the year. We have taken care to ensure that the data we have used was disambiguated and clean by choosing appropriate keywords and performing sanity checks.

\begin{table}
\label{movie_data}
\begin{center}
\begin{tabular}{|c|c|}
\hline
Movie & Release Date\\
\hline
Armored & 2009-12-04\\
\hline
Avatar & 2009-12-18\\
\hline
The Blind Side & 2009-11-20\\
\hline
The Book of Eli & 2010-01-15\\
\hline
Daybreakers & 2010-01-08\\
\hline
Dear John & 2010-02-05\\
\hline
Did You Hear About The Morgans & 2009-12-18\\
\hline
Edge Of Darkness & 2010-01-29\\
\hline
Extraordinary Measures & 2010-01-22\\
\hline
From Paris With Love & 2010-02-05\\
\hline
The Imaginarium of Dr Parnassus & 2010-01-08\\
\hline
Invictus & 2009-12-11\\
\hline
Leap Year & 2010-01-08\\
\hline
Legion & 2010-01-22\\
\hline
Twilight : New Moon & 2009-11-20\\
\hline
Pirate Radio & 2009-11-13\\
\hline
Princess And The Frog & 2009-12-11\\
\hline
Sherlock Holmes & 2009-12-25\\
\hline
Spy Next Door & 2010-01-15\\
\hline
The Crazies & 2010-02-26\\
\hline
Tooth Fairy & 2010-01-22\\
\hline
Transylmania & 2009-12-04\\
\hline
When In Rome & 2010-01-29\\
\hline
Youth In Revolt & 2010-01-08\\
\hline
\end{tabular}
\end{center}
\caption{Names and release dates for the movies we considered in our analysis.}
\end{table}
\begin{figure}[h]
\begin{center}
\includegraphics[height=2in,width=3.5in]{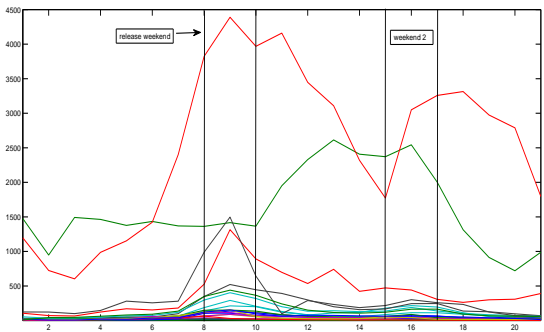}
\caption{Time-series of tweets over the critical period for different movies.}
\label{tweet_trends}
\end{center}
\end{figure}
The total data over the critical period for the 24 movies we considered includes 2.89 million 
tweets from 1.2 million users. 

Fig~\ref{tweet_trends} shows the timeseries trend in the number of tweets for movies over the critical period.  
We can observe 
that the busiest time for a movie is around the time it is released, following which the chatter invariably fades. 
The box-office revenue follows a similar trend with the opening weekend generally providing the most revenue for a 
movie. 

\begin{figure}
\begin{center}
\includegraphics[height=2in,width=3.5in]{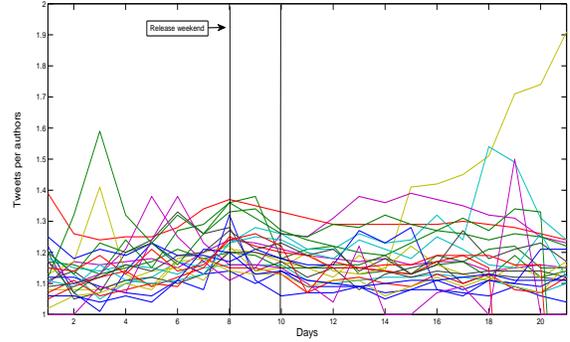}
\caption{Number of tweets per unique authors for different movies }
\label{authts}
\end{center}
\end{figure}
\begin{figure}[h]
\begin{center}
\includegraphics[height=2in,width=3.5in]{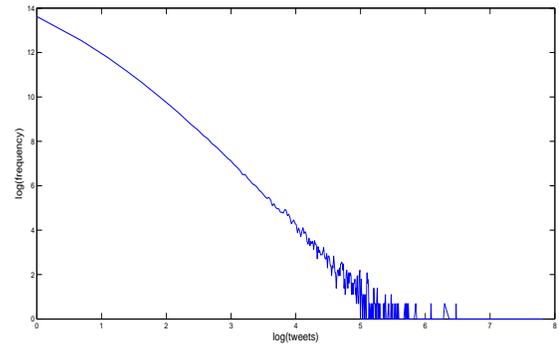}
\caption{Log distribution of authors and tweets. }
\label{authdist2}
\end{center}
\end{figure}
\begin{figure}[h]
\begin{center}
\includegraphics[height=2in,width=3.5in]{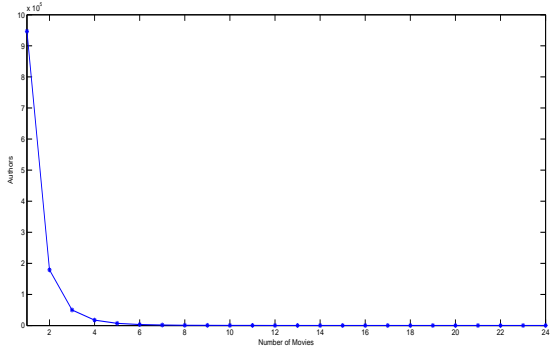}
\caption{Distribution of total authors and the movies they comment on.}
\label{auth_movdist}
\end{center}
\end{figure}

Fig~\ref{authts} shows how the number of tweets per unique author changes over time. We find that this ratio remains fairly consistent with a value between 1 and 1.5 across the critical period. 
Fig~\ref{authdist2} displays the distribution of tweets by different authors over the critical period. The X-axis shows the number of tweets in the log scale, while the Y-axis represents the corresponding frequency of authors in the log scale. We can observe that it is close to a Zipfian distribution, with a few authors generating a large number of tweets. This is consistent with observed behavior from other networks~\cite{Wu2009}.
Next, we examine the distribution of authors over different movies.
Fig~\ref{auth_movdist} shows the distribution of authors and the number of movies they comment on. 
Once again we find a power-law curve, with a majority of the authors talking about only a few movies.

\section{Attention and Popularity}

We are interested in studying how attention and popularity are generated for movies on Twitter, and the effects of 
this attention on the  real-world performance of the movies considered. 

\subsection{Pre-release Attention:} 

Prior to the release of a movie, media companies and and producers generate promotional information in the form of 
trailer videos, news, blogs and photos. We expect the tweets for movies before the time of their release to consist primarily 
of such promotional campaigns, geared to promote word-of-mouth cascades. On Twitter, this can be characterized by tweets referring to particular urls (photos, trailers and other promotional material) as well as retweets, which involve users forwarding tweet posts to everyone in their friend-list. Both these forms of tweets are important to disseminate information regarding movies being released.

First, we examine the distribution of such tweets for different movies, following which we examine their correlation with 
the performance of the movies.

\begin{table}
\label{urlrt}
\begin{center}
\begin{tabular}{|c|c|c|c|}
\hline
Features & Week 0 & Week 1 & Week 2\\
\hline
url & 39.5 & 25.5 & 22.5\\
\hline
retweet & 12.1 & 12.1 & 11.66\\
\hline
\end{tabular}
\end{center}
\caption{Url and retweet percentages for critical week}
\end{table}
\begin{figure}[h]
\begin{center}
\includegraphics[height=2in,width=3.5in]{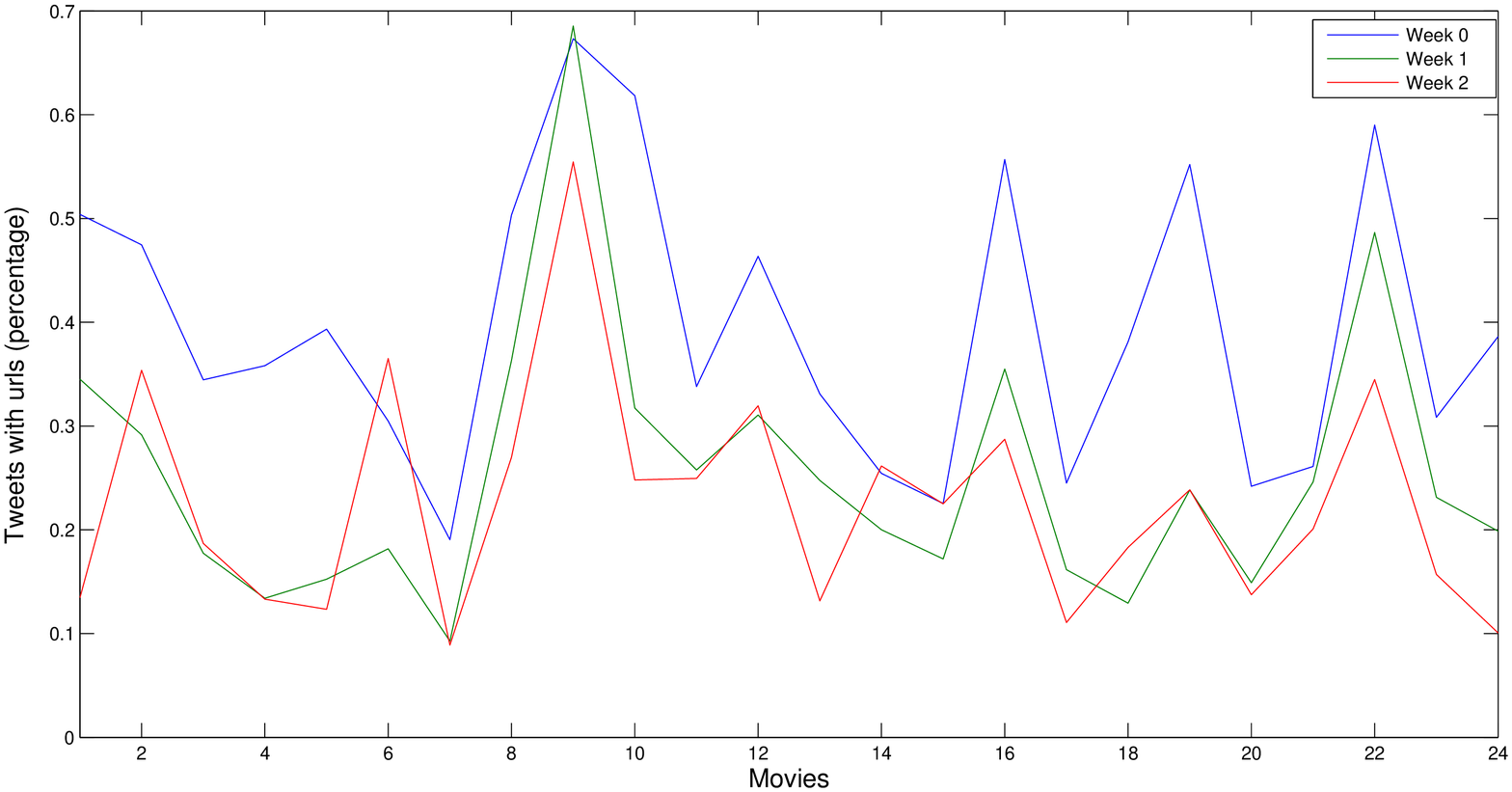}
\caption{Percentages of urls in tweets for different movies. }
\label{authdist}
\end{center}
\end{figure}
\begin{table}
\label{urlrt2}
\begin{center}
\begin{tabular}{|c|c|c|}
\hline
Features & Correlation & $R^{2}$\\
\hline
url & 0.64 & 0.39\\
\hline
retweet & 0.5 & 0.20\\
\hline
\end{tabular}
\end{center}
\caption{Correlation and $R^{2}$ values for urls and retweets before release.}
\end{table}
Table 2 shows the percentages of urls and retweets in the tweets over the critical period for movies. We can observe that there is a greater percentage of tweets containing urls in the week prior to release than afterwards. This is consistent with our expectation. In the case of retweets, we find the values to be similar across the 3 weeks considered. In all, we found the retweets to be a significant minority of the tweets on movies. One reason for this could be that people tend to describe their own expectations and experiences, which are not necessarily propaganda.

We want to determine whether movies that have greater publicity, in terms of linked urls on Twitter, perform better in the box office. When we examined the correlation 
between the urls and retweets with the box-office performance, we found the correlation to be moderately positive, as shown in Table 3. However, the adjusted $R^{2}$ value is quite low in both cases, indicating that these features are not very predictive of the relative performance of movies. This result is quite surprising since we would expect promotional material to contribute significantly to a movie's box-office income. 

\subsection{Prediction of first weekend Box-office revenues}
Next, we investigate the power of social media in predicting real-world outcomes. 
Our goal is to observe if the knowledge that can be extracted from the tweets can lead to reasonably accurate 
prediction of future outcomes in the real world.

The problem that we wish to tackle can be framed as follows. {\it Using the tweets referring to movies prior 
to their release, can we accurately predict the box-office revenue generated by the movie in its opening 
weekend?}

\begin{table}
\label{res}
\begin{center}
\begin{tabular}{|c|c|c|}
\hline
Features & Adjusted $R^{2}$ & p-value\\
\hline
Avg Tweet-rate & 0.80 & 3.65e-09\\
\hline 
Tweet-rate timeseries & 0.93 & 5.279e-09\\
\hline
Tweet-rate timeseries + thcnt & {\bf 0.973} & 9.14e-12\\
\hline
HSX timeseries + thcnt & 0.965 & 1.030e-10\\  
\hline
\end{tabular}
\end{center}
\caption{Coefficient of Determination ($R^{2}$) values using different predictors for movie box-office revenue for the first weekend.}
\end{table}
\begin{figure}[h]
\begin{center}
\includegraphics[height=2in,width=3.5in]{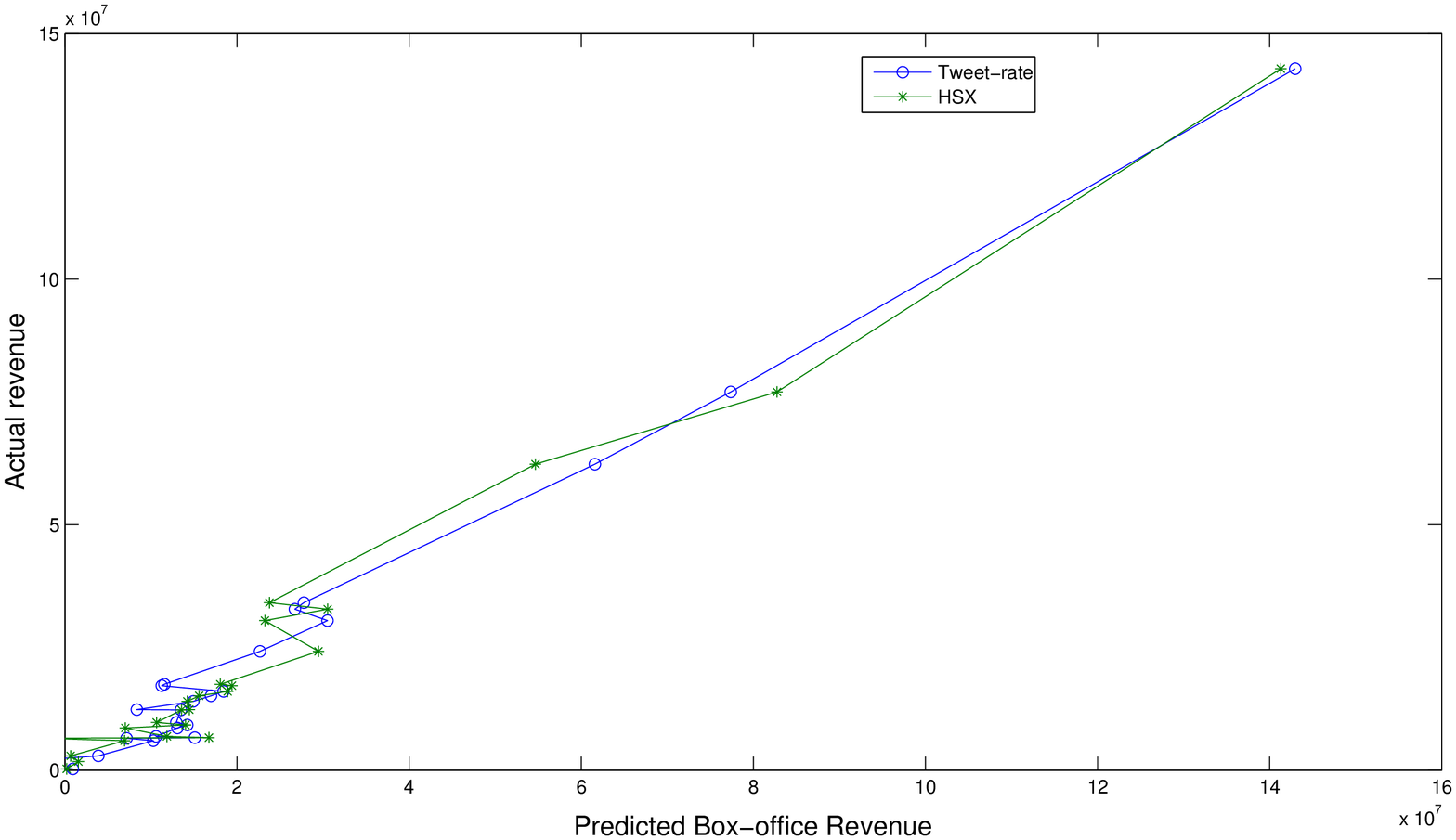}
\caption{Predicted vs Actual box office scores using tweet-rate and HSX predictors}
\label{pred}
\end{center}
\end{figure}

To use a quantifiable measure on the tweets, we define the {\bf tweet-rate}, as the {\it number of tweets referring to a particular movie per 
hour}.
\begin{equation}
Tweet-rate(mov) = \frac{|tweets(mov)|}{|Time\ (in\ hours)|}
\end{equation}

Our initial analysis of the correlation of the average tweet-rate with the box-office gross for the 24 movies considered showed a strong positive correlation, with a correlation coefficient value of 0.90. This suggests a strong linear relationship among the variables considered. Accordingly, we constructed a linear regression model using least squares of the average of all tweets for the 24 movies considered over the week {\it prior to their release}. We obtained an adjusted $R^{2}$ value of 0.80 with a p-value of $3.65e-09***$, where the '***' shows significance at 0.001, indicating a very strong predictive relationship. Notice that this performance was achieved using only one variable (the average tweet rate).  
To evaluate our predictions, we employed real box-office revenue information, extracted from the Box Office Mojo website~\footnote{http://boxofficemojo.com}. 

The movie $Transylmania$ that opened on Dec 4th had easily the lowest tweet-rates of all movies considered. 
For the week prior to its release, it received on an average 2.75 tweets per hour. 
As a result of this lack of attention, the movie captured the record for the lowest-grossing opening for a movie playing 
at over 1,000 sites, making only \$263,941 in its opening weekend, and was subsequently pulled 
from theaters at the end of the second week.
On the other end of the spectrum, two movies that made big splashes in their opening weekends, {\it Twilight:New Moon} (making 142M)
and {\it Avatar}(making 77M) had, for their pre-release week, averages of 1365.8 and 1212.8 tweets per hour respectively. This once again illustrates the importance of attention in social media.

Next, we performed a linear regression of the time series values of the tweet-rate for the 7 days before the release. We used 7 variables each corresponding to the tweet-rate for a particular day. An additional variable we used was the number of theaters the movies were released in, $thcnt$. The results of the regression experiments are shown in Table 4. Note that, in all cases, we are using only data available prior to the release to predict box-office for the opening weekend.\\

{\bf \noindent Comparison with HSX:}\\
To compare with our tweet-based model, we used the Hollywood Stock Exchange index.
The fact that artificial online markets such as the Foresight Exchange and the Hollywood Stock 
Exchange are good indicators of future outcomes has been shown previously~\cite{Pennock2001,Chen2003}. 
The prices in these markets have been shown to have 
strong correlations with observed outcome frequencies. In the case of movies, the Hollywood Stock Exchange (http://www.hsx.com/), 
is a popular play-money market, where the prices for movie stocks can accurately predict real box office results.
Hence, to compare with our tweet-rate predictor, we considered regression on the movie stock prices from the 
Hollywood Stock Exchange, which can be considered the gold standard~\cite{Pennock2001}. 

From the results in Table 4, it can be seen that our regression model built from social media provides an accurate prediction of movie performances at the box office. Furthermore, the model built using the tweet rate timeseries {\it outperforms} the HSX-based model.
\begin{table}
\label{week}
\begin{center}
\begin{tabular}{|c|c|c|}
\hline
Predictor & AMAPE & Score\\
\hline
$Reg_{nobudget}$+$nReg_{1w}$ & 3.82 & 96.81\\
\hline
Avg Tweet-rate + thcnt & 1.22 & 98.77\\
\hline
Tweet-rate Timeseries + thcnt & {\bf 0.56} & {\bf 99.43}\\
\hline
\end{tabular}
\end{center}
\caption{AMAPE and Score value comparison with earlier work.}
\end{table}
The graph outlining the predicted and actual values of this model is also shown in Fig~\ref{pred}, outlining the utility of harvesting social media.\\

{\bf \noindent Comparison with News-based Prediction:}\\
In earlier work, Zhang and others~\cite{Zhang2009} have developed a news-based model for predicting movie revenue. 
The best-performing method in the aforementioned work is the combined model obtained by using predictors from IMDB and news. The corresponding $R^{2}$ value for this combined model is 0.788, which is far lower than the ones obtained by our predictors. We computed the AMAPE  (Adjusted Mean Absolute Percentage/Relative Error) measure, that the authors use, for our data. The comparative values are shown in Table 5. We can observe that our values are far better than the ones reported in the earlier work. Note however, that since historical information on tweets are not available, we were able to use data on only the movies we have collected, while the authors in the earlier paper have used a larger database of movies for their analysis.

\subsection{Predicting HSX prices}

Given that social media can accurately predict box office results, we also tested their efficacy at forecasting the stock prices of the HSX index. At the end of the first weekend, the Hollywood stock exchange adjusts the price for a movie stock to 
reflect the actual box office gross. If the movie does not perform well, the price goes down and vice versa. We conducted an experiment to see if we could predict the price of the HSX movie stock 
at the end of the opening weekend for the movies we have considered. We used the historical HSX prices as well as the tweet-rates, individually, for the week prior to the release as predictive variables. The response variable was the adjusted price of the stock. We also used the theater count as a predictor in both cases, as before.
The results are summarized in Table 6.  As is apparent, the tweet-rate proves to be {\it significantly better} at predicting the actual values than the historical HSX 
prices. This 
again illustrates the power of the buzz from social media.
\begin{table*}
\label{hsx}
\begin{center}
\begin{tabular}{|c|c|c|}
\hline
Predictor & Adjusted $R^{2}$ & $p-value$ \\
\hline
HSX timeseries + thcnt & 0.95 & 4.495e-10\\
\hline
Tweet-rate timeseries + thnt & {\bf 0.97} & 2.379e-11\\
\hline
\end{tabular}
\end{center}
\caption{Prediction of HSX end of opening weekend price.}
\end{table*}

\subsection{Predicting revenues for all movies for a given weekend}
Until now, we have considered the problem of predicting opening weekend revenue for movies. Given the success of the regression model, we now attempt to predict revenue for all movies over a particular weekend. The Hollywood Stock Exchange de-lists movie stocks after 4 weeks of release, which means that there is no timeseries available for movies after 4 weeks. In the case of tweets, people continue to discuss movies long after they 
are released. Hence, we attempt to use the timeseries of tweet-rate, over 7 days before the weekend, to predict the box-office revenue for that particular weekend.
Table 7 shows the results for 3 weekends in January and 1 in February. Note, that there were movies that were two months old 
in consideration for this experiment.
Apart from the time series, we used two additional variables - the theater count and the number of weeks the movie has been 
released. We used the coefficient of determination (adjusted $R^2$) to evaluate the regression models. From Table 7, we find that the tweets continue to be 
good predictors even in this case, with an adjusted $R^{2}$ consistently greater than $0.90$.
\begin{table}
\label{week}
\begin{center}
\begin{tabular}{|c|c|}
\hline
Weekend & Adjusted $R^{2}$ \\
\hline
Jan 15-17 & 0.92\\
\hline
Jan 22-24 & 0.97\\
\hline
Jan 29-31 & 0.92\\  
\hline
Feb 05-07 & 0.95\\
\hline
\end{tabular}
\end{center}
\caption{Coefficient of Determination ($R^{2}$) values using tweet-rate timeseries for different weekends}
\end{table}
\begin{table*}
\label{2ndweek}
\begin{center}
\begin{tabular}{|c|c|c|}
\hline
Predictor & Adjusted $R^{2}$ & $p-value$ \\
\hline
Avg Tweet-rate & 0.79 & 8.39e-09\\
\hline
Avg Tweet-rate + thcnt & 0.83 & 7.93e-09\\
\hline
Avg Tweet-rate + PNratio & 0.92 & 4.31e-12\\
\hline
\hline
Tweet-rate timeseries & 0.84 & 4.18e-06\\
\hline
Tweet-rate timeseries + thcnt & 0.863 & 3.64e-06\\
\hline
Tweet-rate timeseries + PNratio & {\bf 0.94} & 1.84e-08\\
\hline
\end{tabular}
\end{center}
\caption{Prediction of second weekend box-office gross}
\end{table*}
The results have shown that the buzz from social media can be accurate indicators of future outcomes. The fact that a simple linear regression model considering only 
the rate of tweets 
on movies can perform better than artificial money markets, illustrates the power of social media. 

\section{Sentiment Analysis}
Next, we would like to investigate the importance of sentiments in predicting future outcomes. We have seen how efficient 
the attention can be in predicting opening weekend box-office values for movies. Hence we consider the problem of 
utilizing the sentiments prevalent in the discussion for forecasting.

Sentiment analysis is a well-studied problem in linguistics and machine learning, with different classifiers and language models employed in earlier work~\cite{Pang2008,Godbole2007}. It 
is common to express this as a classification problem where a given text needs to be labeled as $Positive$, $Negative$ or $Neutral$.
Here, we constructed a sentiment analysis classifier using the LingPipe linguistic analysis package~\footnote{http://www.alias-i.com/lingpipe} which provides a set of open-source java libraries for natural language processing
tasks. We used the DynamicLMClassifier which is a language model classifier that accepts training events of categorized character sequences. Training is based on a multivariate estimator for the category distribution and dynamic language models for the per-category character sequence estimators. To obtain labeled training data for the classifier, we utilized workers from the Amazon Mechanical Turk~\footnote{https://www.mturk.com/}. It has been shown that manual labeling from Amazon Turk can correlate well with experts~\cite{Snow2008}. We used thousands of workers to assign sentiments for a large random sample of tweets,  ensuring that each tweet was labeled by three different people. We used only samples for which the vote was unanimous as training data. 
\begin{figure}[h]
\begin{center}
\includegraphics[height=2in,width=3.5in]{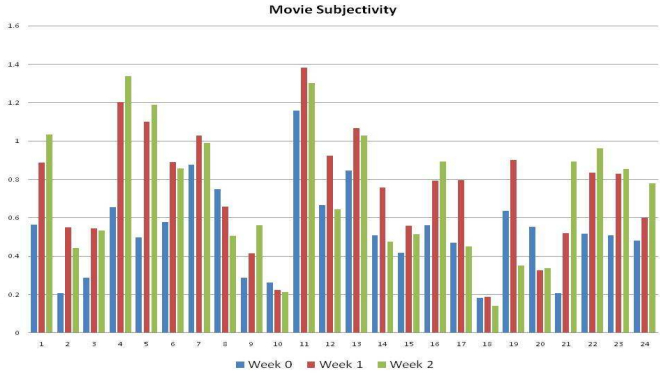}
\caption{Movie Subjectivity values}
\label{subj}
\end{center}
\end{figure}
\begin{figure}[h]
\begin{center}
\includegraphics[height=2in,width=3.5in]{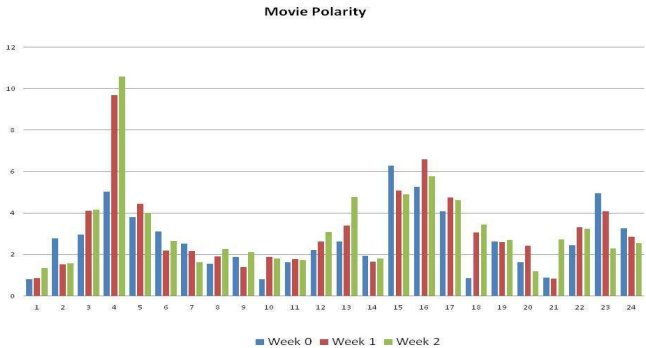}
\caption{Movie Polarity values}
\label{pol}
\end{center}
\end{figure}
The samples were initially preprocessed in the following ways:
\begin{itemize}
\item Elimination of stop-words 
\item Elimination of all special characters except exclamation marks which were replaced by $<EX>$ and question marks ($<QM>$)
\item Removal of urls and user-ids
\item Replacing the movie title with $<MOV>$
\end{itemize}
We used the pre-processed samples to train the classifier using an n-gram model. We chose n to be 8 in our experiments. The classifier was trained to predict three classes - Positive, Negative and Neutral. When we tested on the training-set with cross-validation, we obtained an accuracy of 98\%. 
We then used the trained classifier to predict the sentiments for all the tweets in the critical period for all the movies considered.

\subsection{Subjectivity}

Our expectation is that there would 
be more value for sentiments after the movie has released, than before. We expect tweets prior to the release to be mostly anticipatory 
and stronger positive/negative tweets to be disseminated later following the release. 
Positive sentiments following the release can be considered as recommendations by people who have seen the movie, and are likely to influence others from watching the same movie.
To capture the subjectivity, we defined a measure as follows.
\begin{equation}
Subjectivity = \frac{|Positive\ and\ Negative\ Tweets|}{|Neutral\ Tweets|}
\end{equation}
When we computed the subjectivity values for all the movies, we observed that our hypothesis was true. 
There were more sentiments discovered in tweets for the weeks after release, than in the pre-release week. Fig~\ref{subj} shows the ratio of subjective to objective tweets for all the movies over the three weeks. We can observe that for most of the movies, the subjectivity increases after release.

\subsection{Polarity}
To quantify the sentiments for a movie, we measured the ratio of positive to negative tweets. A movie that has far more positive than negative tweets is likely to be successful. 
\begin{equation}
PNratio = \frac{|Tweets\ with\ Positive\ Sentiment|}{|Tweets\ with\ Negative\ Sentiment|}
\end{equation}

Fig~\ref{pol} shows the polarity values for the movies considered in the critical period. We find that there are more positive 
sentiments than negative in the tweets for almost all the movies. The movie with the enormous increase in positive sentiment after release is {\it The Blind Side} (5.02 to 9.65). The movie had a lukewarm opening weekend sales (34M) but then boomed in the next week (40.1M), owing largely to positive sentiment. 
The movie {\it New Moon} had the opposite effect. It released in the same weekend as {\it Blind Side} and had a great first weekend but its polarity reduced (6.29 to 5), as did its box-office revenue (142M to 42M) in the following week.

\begin{table}
\begin{center}
\begin{tabular}{|c|c|}
\hline
Variable & $p-value$ \\
\hline
(Intercept) & 0.542    \\
\hline
Avg Tweet-rate & 2.05e-11 (***)\\
\hline
PNRatio & 9.43e-06 (***) \\
\hline
\end{tabular}
\end{center}
\caption{Regression using the average tweet-rate and the polarity (PNRatio). The significance level (*:0.05, **: 0.01, ***: 0.001) is also shown.}
\end{table}
Considering that the polarity measure captured some variance in the revenues, we examine the utility of the sentiments in predicting box-office sales. 
In this case, we considered the second weekend revenue, since we have seen subjectivity increasing after release. We use linear regression on the revenue as before, using the tweet-rate and the PNratio as an additional variable.
The results of our regression experiments are shown in Table 8.  We find that the sentiments do provide improvements, although they are not as important as the rate of tweets themselves. The tweet-rate has close to the same predictive power in the second week as the first. Adding the sentiments, as an additional variable, to the regression equation improved the prediction to 0.92 while used with the average tweet-rate, and 0.94 with the tweet-rate timeseries. Table 9 shows the regression p-values using the average tweet rate and the sentiments. We can observe that the coefficients are highly significant in both cases.

\section{Conclusion}

In this article, we have shown how social media can be utilized to forecast future outcomes. Specifically, using the rate of chatter from almost 3 million tweets from the popular site Twitter, we constructed a linear regression model for predicting box-office revenues of movies in advance of their release. We then showed that the results outperformed in accuracy those of the Hollywood Stock Exchange and that there is a strong correlation between the amount of attention a given topic has (in this case a forthcoming movie) and its ranking in the future. We also analyzed the sentiments present in tweets and demonstrated their efficacy at improving predictions after a movie has released.

While in this study we focused on the problem of predicting box office revenues of movies for the sake of having a clear metric of comparison with other methods, this method can be extended to a large panoply of topics, ranging from the future rating of products to agenda setting and election outcomes. At a deeper level, this work shows how social media expresses a collective wisdom which, when properly tapped, can yield an extremely powerful and accurate indicator of future outcomes.

\section{Appendix: General Prediction Model for Social Media}

Although we focused on movie revenue prediction in this paper, the method that we advocate can be extended to other products of consumer interest. 

We can generalize our model for predicting the revenue of a product using social media as follows. We begin with data 
collected regarding the product over time, in the form of reviews, user comments and blogs. Collecting the data over time is important as it can measure the rate of chatter effectively.
The data can then be used to fit a linear regression model using least squares. The parameters of the model include:
\begin{itemize}
\item $A$ : rate of attention seeking
\item $P$ : polarity of sentiments and reviews
\item $D$ : distribution parameter
\end{itemize}
Let $y$ denote the revenue to be predicted and $\epsilon$ the error. The linear regression model can be expressed as :
\begin{equation}
y = \beta_{a}*A +\beta_{p}*P +\beta_{d}*D + \epsilon
\end{equation}
where the $\beta$ values correspond to the regression coefficients.
The attention parameter captures the buzz around the product in social media. In this article, we showed how the rate of tweets on Twitter can capture attention on movies accurately. We found this coefficient to be the most significant in our experiments.
The polarity parameter relates to the opinions and views that are disseminated in social media. We observed that this gains importance after the movie has been released and adds to the accuracy of the predictions. 
In the case of movies, the distribution parameter is the number of theaters a particular movie is released in. In the case of other products, it can reflect their availability in the market.

\section{Acknowledgement}
This material is based upon work supported by the National Science Foundation under Grant $\#$ 0937060 to the Computing Research Association for the CIFellows Project.
\bibliographystyle{IEEETran}

\end{document}